\documentclass[conference]{IEEEtran}
\IEEEoverridecommandlockouts
\usepackage{framed}

\usepackage[sort&compress,square,comma,numbers]{natbib}
\usepackage[acronym,shortcuts,nomain,style=super,nonumberlist,nopostdot]{glossaries} 
\usepackage{soul}
\usepackage{multirow}
\usepackage{booktabs}
\usepackage{cleveref}
\usepackage{todonotes}
\usepackage{subcaption}
\usepackage{wrapfig}
\usepackage{progressbar}
\usepackage{url}
\usepackage{comment}
\usepackage{subcaption}
\usepackage{stfloats}

\crefformat{figure}{Fig.~#2#1#3} 

\newacronym{iot}{IoT}{Internet of Things}
\newacronym{ai}{AI}{Artificial Intelligence}
\newacronym{xai}{XAI}{Explainable AI}
\newacronym{pca}{PCA}{Principal Component Analysis}
\newacronym{pc}{PC}{Principal Component}
\newacronym{dss}{DSS}{Decision Support System}

\newcommand{\pps}{Persona Perception Scale}

\newcommand{\varGen}{d$_{\text{Gender}}$}
\newcommand{\varAge}{d$_{\text{Age}}$}
\newcommand{\varFarmSize}{d$_{\text{Cows}}$}
\newcommand{\varEdu}{d$_{\text{Edu}}$}
\newcommand{\varSysConf}{d$_{\text{SysConf}}$}
\newcommand{\varSysFreq}{d$_{\text{SysFreq}}$}
\newcommand{\varSysDur}{d$_{\text{SysDur}}$}
\newcommand{\varFarmDur}{d$_{\text{FarmDur}}$}

\newcommand{\varFeedD}{e$_{\text{FeedTech}}$}
\newcommand{\varBreedD}{e$_{\text{BreedTech}}$}
\newcommand{\varHealthD}{e$_{\text{HealthTech}}$}
\newcommand{\varFeedP}{e$_{\text{FeedPriv}}$}
\newcommand{\varBreedP}{e$_{\text{BreedPriv}}$}
\newcommand{\varHealthP}{e$_{\text{HealthPriv}}$}

\newcommand{\rqu}[1]{\textbf{RQ$_{#1}$}}

\graphicspath{{images/}}

\def\BibTeX{{\rm B\kern-.05em{\sc i\kern-.025em b}\kern-.08em
    T\kern-.1667em\lower.7ex\hbox{E}\kern-.125emX}}
\begin{document}

\makeatletter 
\newcommand{\linebreakand}{%
  \end{@IEEEauthorhalign}
  \hfill\mbox{}\par
  \mbox{}\hfill\begin{@IEEEauthorhalign}
}
\makeatother 

\title{Explainability Needs in Agriculture:\\Exploring Dairy Farmers' User Personas}

\author{\IEEEauthorblockN{Mengisti Berihu Girmay}
    \IEEEauthorblockA{
    University of Kaiserslautern-Landau\\
    Kaiserslautern, Germany \\
    mengisti.berihu@rptu.de}
\and
\and
\IEEEauthorblockN{Jakob Droste}
    \IEEEauthorblockA{Software Engineering Group\\
    Leibniz University Hannover\\
    Hannover, Germany \\
    jakob.droste@inf.uni-hannover.de}
\and
\IEEEauthorblockN{Hannah Deters}
    \IEEEauthorblockA{Software Engineering Group\\
    Leibniz University Hannover\\
    Hannover, Germany \\
    hannah.deters@inf.uni-hannover.de}
\and
\linebreakand 
\IEEEauthorblockN{Joerg Doerr}
    \IEEEauthorblockA{University of Kaiserslautern-Landau\\and Fraunhofer IESE\\
    Kaiserslautern, Germany \\
    joerg.doerr@iese.fraunhofer.de}
}

\maketitle

\begin{abstract}

\ac{ai} promises new opportunities across many domains, including agriculture.
However, the adoption of AI systems in this sector faces several challenges.
System complexity can impede trust, as farmers' livelihoods depend on their decision-making and they may reject opaque or hard-to-understand recommendations.
Data privacy concerns also pose a barrier, especially when farmers lack clarity regarding who can access their data and for what purposes.

This paper examines dairy farmers’ explainability requirements for technical recommendations and data privacy, along with the influence of socio-demographic factors.
Based on a mixed-methods study involving 40 German dairy farmers, we identify five user personas through k-means clustering. 
Our findings reveal varying requirements, with some farmers preferring little detail while others seek full transparency across different aspects.
Age, technology experience, and confidence in using digital systems were found to correlate with these explainability requirements.
The resulting user personas offer practical guidance for requirements engineers aiming to tailor digital systems more effectively to the diverse requirements of farmers.

\end{abstract}


\glsresetall

\section{Introduction}

\ac{ai} is advancing in many areas and promises new opportunities to increase productivity and efficiency.
Combined with the increasing availability of low-cost sensors, \ac{ai} has also sparked renewed interest in precision agriculture, an approach that uses data-driven insights to optimize farm management and resource use.
AI is also finding its way into livestock farming, where  pattern recognition can help detect animal health issues, e.g., lameness using image data or respiratory diseases such as coughing through sound analysis.
This not only promises to improve animal welfare, but also to enable early interventions, reducing the need for antibiotics and helping prevent larger disease outbreaks \cite{monteiro2021precision}.

However, despite the large potential and many use cases, practical adoption of AI systems is still lacking.
AI systems can raise concerns for farmers, particularly when the reasoning behind decisions or predictions is unclear.
Unlike rule-based systems of conventional computer programming, in which the decision-making logic is predefined and transparent, AI models tend to operate as “black boxes”, making it difficult or impossible for users (and often even developers) to understand exactly how certain conclusions are reached.

This “black box” issue presents a particular trust challenge for AI in agriculture on two interrelated levels.
First, when farmers rely on AI systems for technical guidance but cannot comprehend the rationale behind recommendations, they may disregard or reject them entirely --- given that their livelihood depends on their decisions.
A lack of transparency, high system complexity, and fears of error or loss of control can significantly undermine trust in such systems \cite{dorr2022handbook}.
Second, many farmers are concerned about data privacy and fear that their data could be accessed by authorities or monetized by companies without their consent \cite{linsner2021role}.
If AI systems and their data handling practices are perceived as opaque or externally imposed, they may be met with resistance and ultimately go unused.
Thus, AI should not only be intuitive for end users but also transparent and adapted to their requirements to support user acceptance and adoption.

Research in both agricultural \cite{girmay2025perspectives} and non-agricultural domains \cite{chazette2022explainable, kohl2019explainability} has shown that integrating explanations into AI may increase user trust and promote adoption.
However, explanations can also have a negative impact on usability if they are not helpful and rather burden the user (e.g., by cluttering the user interface) \cite{chazette2022explainable}.
For this reason, explanations should be well adapted to user requirements.

An established method for capturing the requirements of stakeholder groups is the development of user personas.
To create them, users with similar requirements are clustered and examined for similarities, e.g., with regard to their sociodemographic characteristics.
In this way, user personas systematically represent the needs and human values of different user groups and help to ensure that users who might otherwise be overlooked are also addressed.
By making end users' requirements more tangible, personas enable requirements engineers better empathize with them and guide system development in a way that is ethically sound and inclusive  \cite{droste2023designing}.

Building on this, we aim to explore the requirements of dairy farmers in terms of explainability.
To build a data foundation, we surveyed 40~dairy farmers in Germany and interviewed eight of them for deeper insights into their responses and views on system explanations.
The survey focused on their requirements regarding explanations, including the desired level of detail in technical and data privacy explanations across scenarios related to feeding, breeding, and animal health.
We clustered their feedback into five user personas using k-means.
Each persona represents a typical farmer, combining sociodemographic traits with related explainability requirements.
These personas are intended to support requirements engineers by providing a more tangible picture of the target group and their various explanatory needs.

The remainder of the paper is structured as follows.
\Cref{sec:background-and-related-work} summarizes relevant background information and related work.
In \Cref{sec:research-design-and-methodology}, we present our research design and methodology.
\Cref{sec:results} presents the findings of our work. In \Cref{sec:discussion}, we discuss our results.
\Cref{sec:conclusion} concludes the paper with a brief summary and an outlook on future work.

\section{Background and Related Work}
\label{sec:background-and-related-work}

In the following, we compile relevant background information and related work.

\subsection{Explainability}

Explainability commonly refers to the extent to which systems that provide assessments or recommendations make their reasoning transparent~\cite{miller2019explanation}.
The concept relates to how a system explains its decisions rather than what it does.
A key aspect in this context is identifying the intended audience.
For example, while a machine learning engineer may look for insights into model internals, end users are more likely to look for support in interpreting the results~\cite{preece2018stakeholders}.

Explainability has gained significant attention in recent years, especially in the context of AI~\cite{adadi2018peeking, arrieta2020explainable}.
Applications include healthcare~\cite{saraswat2022explainable}, automotive~\cite{atakishiyev2024explainable}, and finance~\cite{kuiper2022exploring}.
As systems grow more complex, the concept is also gaining importance beyond AI, including in areas like data privacy~\cite{brunotte2023context}.
Gaining a better understanding of user behavior and when and how detailed explanations users prefer has the potential to build trust and improve adoption~\cite{brunotte2023context, chazette2020explainability, deters2025identifying}. 
To achieve this, explanations must be appropriately tailored, as excessive or poorly timed information can negatively affect user experience~\cite{chazette2020explainability, deters2024x}.
Effective explanations provide enough detail to support understanding without overwhelming or confusing the user~\cite{weatherson2012explanation}.
This suggests a “one-size-fits-all” approach to explainability is not feasible.

In order to better understand what constitutes good explanations, \citet{kohl2019explainability} recommend bringing together a diverse group of representatives from relevant target groups (in our case farmers).
Including participants with varying demographics (e.g., age, gender, and experience) can help to gather feedback from diverse perspectives.
By asking targeted questions about how tailored explanations should ideally look in each individual case, a well-founded understanding of what explainability should encompass can be established \cite{langer2021we}.
In this paper, we combine a quantitative method (survey) with a qualitative approach (interviews) to capture both broad input and in-depth insights into user requirements, following an approach successfully applied in prior research~\cite{chazette2022can, droste2023designing}.


Efforts to improve explainability are also made in agriculture, although the focus is mostly on developers and researchers, with little involvement from end users.
For instance, \citet{shams2024enhancing} integrated an \ac{xai}-based algorithm into a yield estimation system, but did not include farmers in the validation.
Similarly, studies on plant disease detection~\cite{10151399} or plant identification \cite{mengisti2024explainable} are mostly concerned with technical interpretability over user-centered evaluation.
We argue that farmers and their needs should play a more central role, as their satisfaction ultimately determines acceptance and practical adoption.
Studies have shown that adoption can also be influenced by sociodemographic factors such as age, farm size, education, and prior experience with digital systems~\cite{Michels2019}.
We include these factors to examine how they correlate with farmers’ explainability requirements.


\subsection{User Personas}

The concept of user persona studies is a promising way to analyze and group user requirements~\cite{karolita2023use}.
While there is no universal template for defining user personas, they commonly categorize stakeholders based on shared traits~\cite{salminen2020persona}.
In doing so, personas help explore how users interact with systems and what factors, such as sociodemographics, influence their behavior.
Scenarios can further support this by illustrating user behavior in specific contexts~\cite{karolita2023use}.

User personas have been used for requirements engineering in a wide range of areas, including security and data privacy \cite{WEBERDUPREE20181, dupree2016privacy}, as well as explainability \cite{droste2023designing}.
To the best of our knowledge, however, there are no studies in the literature that apply the user persona concept to study explainability requirements in the agricultural domain.
Given the diversity of domains, such as medical, automotive, and agriculture, findings from one area are difficult to generalize to another.
For example, explainability needs in the automotive domain may focus on understanding system alerts and component failures, whereas farmers face different tasks that likely lead to other distinct requirements.

\section{Research Design and Methodology}
\label{sec:research-design-and-methodology}

\begin{figure}[t]
    \centering
    \includegraphics[width=.715\linewidth]{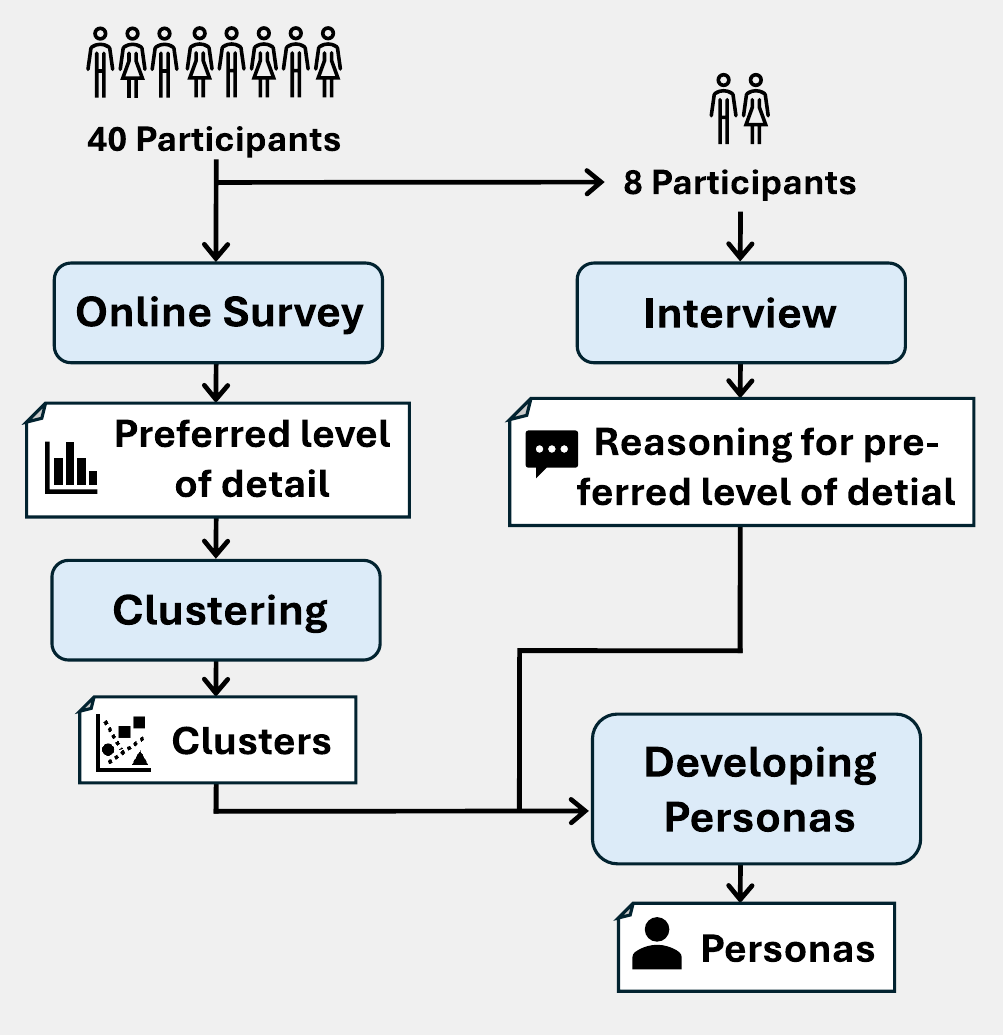}
    \caption{User persona creation process}
    \label{fig:methodology}
\end{figure}

The primary goal of this work is to explore dairy farmers' explainability requirements, develop user personas, and analyze correlations with sociodemographic factors.
In the survey shown to farmers, we asked about their preferred level of detail in technical explanations (assessments or recommendations by the system) and data privacy explanations (clarity on where and how data is stored and processed).
Interview responses were used to derive persona traits and capture farmers’ backgrounds and attitudes.
Both the quantitative and qualitative parts of the study were conducted in parallel.
Our research is guided by the following two questions:

\vspace{3mm}

\begin{tabular}{@{}lp{.82\linewidth}}
\rqu{1}: & Which personas can be identified for dairy farmers, and what are their explainability requirements? \\
\rqu{2}: & Do farmers’ sociodemographic factors correlate with their explainability requirements? \\
\end{tabular}

\subsection{Survey}


We designed the survey based on three scenarios in major decision areas for dairy farmers: feeding, breeding, and health management.
Each scenario presented a hypothetical system assessment (e.g., a cow infected with mastitis) alongside five explanations with varying levels of detail.
Participants chose the level of detail that suited them best, with the order of options randomized to avoid primacy-recency effects~\cite{felfernig2007persuasive}.
To this end, we presented exemplary explanations in order to make the various levels of detail more tangible.
By using concrete examples, participants did not have to interpret abstract labels (such as 'little detail' or 'high detail'), which could have rendered the selection process ambiguous and subjective due to varying understandings.
Similarly, participants chose one of five data privacy explanations ranging from no explanation to highly detailed information, also presented with concrete examples in randomized order.
This yielded six variables, namely the desired level of detail for technical and data privacy explanations for each of the three areas.
The survey further gathered sociodemographic data to enable correlation analysis with the requirements.
The sociodemographic variables included age, gender, farm size, highest educational degree, frequency and duration of digital system use, and years of experience in dairy farming.

\subsection{Interviews}

The interviews were organized informally based on the participants' responses to the survey.
The questions asked did not follow a predefined interview guide.
Instead, each farmer was asked to explain their answers for deeper insight.
Interview lengths varied based on farmers’ availability and ranged from 10 to 25 minutes.
The purpose of the interviews was to obtain qualifying background information on the answers to the survey questions.

\subsection{Data Collection and Analysis}

Farmers were contacted via German farmer associations and personal contacts.
Data collection took place from December 2024 to February 2025.
Eight of the participants agreed to be interviewed in addition to filling out the survey.
The online survey was conducted in German using LimeSurvey~\cite{limesurvey} and later translated into English.
Out of 57 participants, 40 completed the survey in full, and only these complete responses were analyzed. 
The interviewed farmers were contacted based on their availability.

We applied \ac{pca}~\cite{hotelling1933analysis} to reduce the data’s dimensionality.
\ac{pca} transforms data into uncorrelated \acp{pc} that capture the most significant variance.
In doing so, we selected components that together accounted for 95\% of the total variance.

For clustering, we used k-means, an established method for partitioning \ac{pca}-reduced data~\cite{ding2004k}.
The k-means algorithm produces clear, distinct clusters with unambiguous assignments (unlike methods such as DBSCAN~\cite{kanagala2016comparative} where assignments can be ambiguous).
This aspect facilitates deriving user personas.

Using the selected \acp{pc}, we determined the optimal cluster number~(k) to maximize within-cluster similarity and cluster separation.
We applied the elbow method~\cite{cui2020introduction}, which identifies the point where the within-cluster sum of squares (WCSS) decline slows (see \Cref{fig:elbow}), and the silhouette score~\cite{saputra2020effect}, which evaluates cluster quality from +1 (well-separated) to -1 (misclassified) (see \Cref{fig:silhouette}).
Both methods indicated k = 5 as optimal for k-means clustering.

\begin{figure}[h!tb]
    \centering
    \begin{subfigure}{0.45\textwidth}
        \centering
        \includegraphics[width=\textwidth]{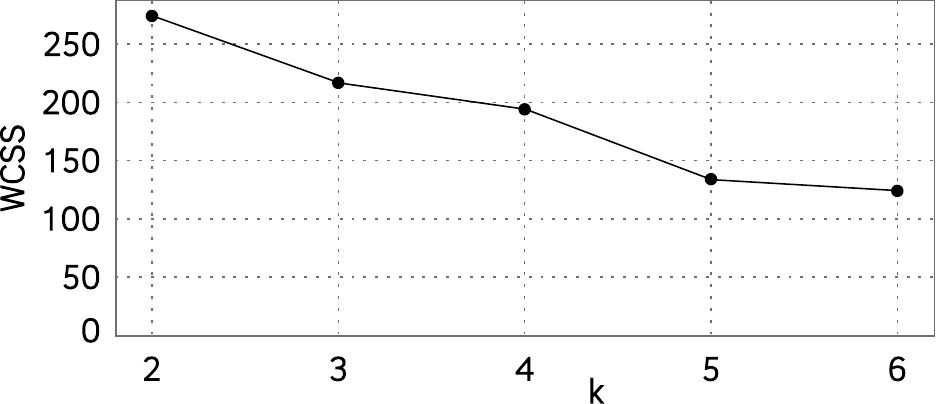}
        \caption{Elbow graph}
        \label{fig:elbow}
    \end{subfigure}
    
    \vspace{3mm}
    
    \begin{subfigure}{0.45\textwidth}
        \centering
        \includegraphics[width=\textwidth]{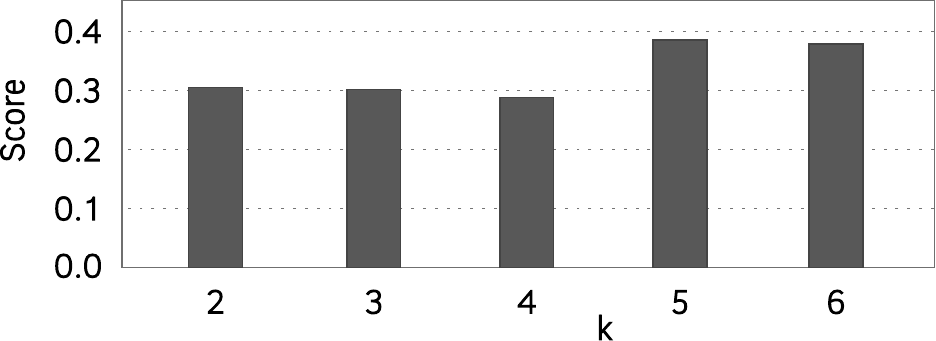}
        \caption{Silhouette scores}
        \label{fig:silhouette}
    \end{subfigure}
    \caption{Determining k for k-means: Elbow and Silhouette}
\end{figure}

To answer \rqu{2}, which explores how sociodemographic factors relate to preferred explanation detail (technical and privacy) across the three scenarios, we tested null hypotheses for each pairwise combination of variables. Pearson correlation \cite{gogtay2017principles} was used for ordinal data, and the Mann-Whitney U test \cite{macfarland2016mann} for nominal data. Relationship direction was determined via r-values (Pearson) and z-values (Mann-Whitney), with significance assessed by p-values. Correlations were considered significant at p~$<$~0.05.

\subsection{Data Availability Statement}

We provide the survey questionnaire with all questions and answer options as supplementary material \cite{girmay_explainability_needs_agriculture_2025}. The raw data from the survey is subject to confidentiality agreements with the participants and cannot be shared publicly.

\section{Findings}
\label{sec:results}

\newcommand{\technicalExplanationNeeds}{Technical details}
\newcommand{\personaFraction}[1]{This persona accounted for #1\% of the participants.}
\newcommand{\personaImage}[1]{\includegraphics[width=.17\textwidth]{#1}}
\newcommand{\personaTitle}[1]{\textit{#1}}
\newcommand{\personaName}[1]{\textbf{#1}} 

\Cref{fig:personas} illustrates the resulting user personas to address \rqu{1}.
The replies from a total of 40 dairy farmers were evaluated, of which 12 were women and 28 men.
Their average age was 39.2 years (range: 22–67, median: 38). Farm sizes varied: three managed up to 30 cows, 17 had 31–70, another 17 had 71–150, 2 had 151–300, and one managed over 300 cows.

Regarding education, three farmers were career changers, seven had agricultural training, ten had attended higher agricultural schools, twelve were certified master farmers, and eight held university degrees in agriculture.
We ranked these educational backgrounds in the given order.
Regarding confidence with digital systems: Three farmers described themselves as less confident, 18 as somewhat confident, ten as confident, and nine as very confident.
Regarding years of experience in dairy farming: One farmer reported up to one year, twelve had 6–10 years, 19 had 11–20 years, and eight had more than 20 years.
Regarding frequency of digital system use: Two farmers reported no use, two used them up to once a month, 14 several times per month, six several times per week, and 16 almost daily.
Regarding duration of digital system use: One farmer had used digital systems for up to one year, 13 for 2–3 years, 14 for 4–5 years, and twelve for more than five years.
Participants were also asked whether dairy farming was their main source of income.
However, due to the lack of variability (37 answered yes, only three answered no), we excluded this parameter from the analysis.



\begin{figure*}[!hbt]
\begin{tabular}{ccccc}
\personaImage{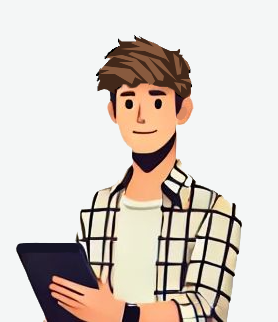} & \personaImage{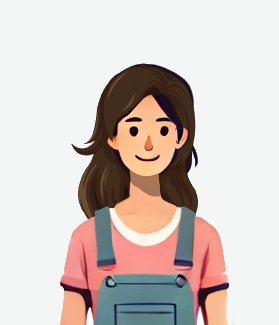} & \personaImage{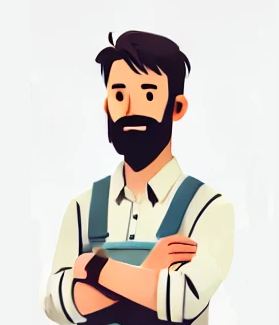} & \personaImage{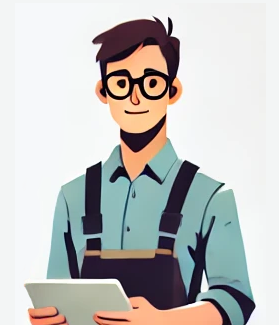} & \personaImage{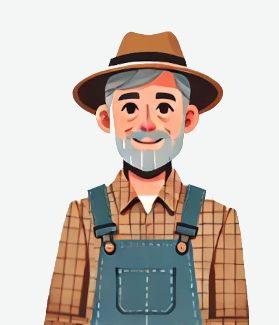} \\
\personaTitle{The Carefree} & \personaTitle{The Detail-Oriented} & \personaTitle{The Minimalist} & \personaTitle{The Pragmatist} & \personaTitle{The Selective} \\
\personaTitle{Power-User} & \personaTitle{Privacy-Skeptic} &  &  & \personaTitle{Privacy-Skeptic} \\[.5em]
\personaName{Tobias Meyer} & \personaName{Anna Weber} & \personaName{Lukas Beck} & \personaName{Michael Fischer} & \personaName{Johann Braun} \\[.5em]
\technicalExplanationNeeds & \technicalExplanationNeeds & \technicalExplanationNeeds & \technicalExplanationNeeds & \technicalExplanationNeeds \\
\progressbar{0.87} & \progressbar{0.72} & \progressbar{0.39} & \progressbar{0.7} & \progressbar{0.46} \\
Privacy details & Privacy details & Privacy details & Privacy details & Privacy details \\
\progressbar{0.4} & \progressbar{0.92} & \progressbar{0.53} & \progressbar{0.6} & \progressbar{1} \\
\end{tabular}
    \caption{Dairy farmer user personas}
    \label{fig:personas}
\end{figure*}

\subsection{User Personas}

The compiled user personas are primarily based on the quantitative survey data, with ordinal variables averaged to define attributes like age and farm size.
Gender was assigned based on cluster majority.
For each user persona, we used the qualitative interview feedback from the corresponding participants to refine the personality and compile an illustrative quote.
Furthermore, we illustrated the desired level of detail for explanations regarding technical aspects (abbreviated as ``technical details'') and data privacy (abbreviated as ``privacy details'').

\subsubsection{The Carefree Power-User}



At 29, Tobias Meyer manages 90 cows and is highly engaged with digital systems.
With six years of dairy farming experience and a degree in agriculture, he confidently uses digital systems daily and has done so for four years.
He values detailed technical explanations to understand system recommendations and optimize his operations.
Data privacy is not a major concern for him, as he generally trusts providers.
``The systems save me time because I don’t have to check everything manually.
But I need to know why I get certain recommendations and what factors led to them.
As for data privacy, I just hope companies handle my data properly''.
\personaFraction{12.5}

\subsubsection{The Detail-Oriented Privacy-Skeptic}


Anna Weber, 39, has run her 50-cow dairy farm for 15 years. Educated at a higher agricultural school, she uses digital systems almost daily and has done so for five years.
While she sees clear benefits in saving time and boosting efficiency, some functionalities are not always intuitive for her and she wants detailed technical explanations.
Data privacy is a major concern for her.
She insists on knowing how and where her data is processed and expects full transparency.
``I find these systems useful, but I need to know exactly what happens to my data.
If a system makes a recommendation, I want to understand why and be able to verify it''.
\personaFraction{25.0}


\subsubsection{The Minimalist}



Lukas Beck, 35, has 10 years of experience in dairy farming and manages 50 cows.
With agricultural training, he is moderately confident using digital systems and has worked with them for five years.
He uses digital systems several times a month, but prefers to rely on his own experience.
He values clear explanations but does not want systems to overwhelm him with excessive detail, both in technical content and data privacy.
``I want to know when animals are having issues or are in heat, but I don’t need long explanations.
Systems should simply work well and explain things only when I ask''.
\personaFraction{17.5}


\subsubsection{The Pragmatist}



Michael Fischer, 44, is a certified master farmer with 17 years of experience, managing 110 cows.
He has been using digital systems multiple times per week for four years and feels confident with them.
He values technical explanations that clarify how and why systems make decisions, but without unnecessary detail.
Regarding data privacy, basic transparency is enough for him and he does not seek in-depth detail.
Michael sees digital systems as useful aids that support (rather than replace) his expertise.
``Explanations are useful if they give me something I can act on. I’m not overly worried about data privacy, but I still expect basic transparency''.
\personaFraction{20.0}


\subsubsection{The Selective Privacy-Skeptic}



Johann Braun, 62, has 20 years of dairy farming experience and manages 70 cows.
Although digital systems were not part of his education, he adopted them confidently four years ago and now uses them weekly.
He relies on them in areas where he feels less confident, like breeding or health management, and expects moderate explanations.
In areas where he is confident, like feeding, he prefers to rely on his own judgment and does not seek explanations.
Data privacy is a major concern and he expects full transparency, mostly due to worries about unauthorized access.
``If I lack experience, explanations are of great help.
But I know my feeding routine in and out and don’t need a system explaining me how to do that.
In other areas, I want to understand why the system suggests something''.
\personaFraction{25.0}

\subsection{Correlation Analysis}


\begin{table}[!b]
    \centering

    \caption{Significant correlations (p~$<$~0.05) between sociodemographic factors (Var.~1) and explanation needs (Var.~2)}
    \label{tab:correlation_results}
    
    \begin{tabular}{c|ll|ll}
        \toprule
        \textbf{Finding} & \textbf{Variable 1} & \textbf{Variable 2} & \textbf{r-value or z-value} & 
        \textbf{p-value} \\
        \midrule
        F$_1$ & \varGen & \varFeedD & z = -2.668 & p = 0.008 \\
        F$_2$ & \varAge & \varBreedP & r = 0.453 & p = 0.003 \\
        F$_3$ & \varAge & \varFeedD & r = -0.340 & p = 0.032 \\
        F$_4$ & \varSysConf & \varBreedP & r = 0.403 & p = 0.010 \\
        F$_5$ & \varSysConf & \varFeedD & r = -0.451 & p = 0.003 \\
        F$_6$ & \varSysFreq & \varFeedD & r = 0.415 & p = 0.008 \\
        F$_7$ & \varSysDur & \varFeedD & r = 0.339 & p = 0.032 \\
        \bottomrule
    \end{tabular}

    \vspace{3mm}

    \begin{tabular}{lp{.8\linewidth}}
        \varGen & Gender (female, male, other) \\
        \varAge & Age group ($\leq$ 25, 26 to 40, 41 to 60, $\geq$ 61) \\
        \varSysConf & Confidence in using digital systems (low to high) \\
        \varSysFreq & Frequency of digital system use (not at all to almost daily) \\
        \varSysDur & Years of digital system use ($\leq$ 1 to $>$ 5 years) \\
        \varFeedD & Desired detail level for feeding-related technical explanations \\
        \varBreedP & Desired detail level for breeding-related privacy explanations \\     
    \end{tabular}    
    
\end{table}

\Cref{tab:correlation_results} shows the results for \textbf{RQ2}, which investigates correlations between farmers' sociodemographic factors and their requirements for explanations.
The table shows only correlations where the null hypothesis was rejected, implying significant correlations (p~$<$~0.05).

The analysis led to the following findings.
F$_1$: In the area of feeding, female farmers preferred more detailed technical explanations than the male participants.
F$_2$ and F$_3$: Farmers of older age showed greater concern about data privacy in breeding and preferred more detailed technical explanations in feeding.
F$_4$ and F$_5$: Farmers with higher confidence in using digital systems desired more detailed data privacy explanations in breeding and less technical detail in feeding.
F$_6$ and F$_7$: Farmers with more frequent or longer-term use of digital systems tend to desire more detailed technical explanations in feeding.

\subsection{Threats to Validity}

Our research faces several validity threats that we discuss below in line with \citet{wohlin2012experimentation}.

\paragraph{Construct Validity}

All participants were dairy farmers and thus relevant stakeholders. Although our survey relied on hypothetical scenarios, we included realistic explanation examples to reduce hypothetical bias.

The quality of user personas depends on how the underlying data is gathered.
Personas based solely on survey data risk being shallow, whereas personas based only on interviews lack quantitative grounding.
We used a mixed-methods approach to combine both methodologies.

Our study focused on text-based explanations, omitting other forms like audio or visuals.
As this work exists in the early stages of explainability research in agriculture, we decided to focus on information content and granularity for the development of our personas.
Future research into presentation forms might yield interesting findings on how farmers prefer their explanations to be provided and presented.

\paragraph{Internal Validity}

Designer bias is a common issue in persona development.
To reduce this, two explainability researchers without ties to digital agriculture independently reviewed the personas.
Nonetheless, some bias may remain.

\paragraph{Conclusion Validity}

With 40 participants, our findings have limited generalizability.
However, given the challenge of recruiting farmers, we consider our sample meaningful.
We do not claim full coverage of all explainability personas among German farmers.

We applied k-means clustering after \ac{pca}, with $k = 5$ supported by elbow and silhouette methods.
The clusters appeared balanced, supporting their use as a persona basis.
Correlation analyses used suitable methods (Pearson's r, Mann-Whitney U), with all findings meeting 5\% significance and indicating moderate correlations, suggesting representativeness.

\paragraph{External Validity}

All participants were German, limiting generalizability across regions. While age diversity was broad, 70\% of participants were male.
However, this aligns with BMEL’s 2020 data \cite{BMEL2023} showing a 36\% share of women among German farmers.

\section{Discussion}
\label{sec:discussion}

User personas, including those developed in our study, can support requirements elicitation and analysis by highlighting the diversity of requirements and preferences within a user group. 
They can help developers gain a shared understanding of their end users and encourage them to consider user types that might otherwise be overlooked by fostering empathy for these user groups. For example, personas may raise awareness of users with strong data privacy concerns who need more transparency and control, or users who prefer minimal explanations and quick overviews, which might prompt developers to offer an option to disable certain explanatory features.

The user personas presented in this paper explicitly refer to explanation needs in agricultural systems, considering both system transparency and user privacy. 
In isolation, explainability and privacy -- and their related user personas -- have already been addressed by previous works. Droste et al.~\cite{droste2023designing} created personas for explanation needs in everyday software systems. They referred, among other things, to demographic factors such as age and occupational field, and clustered software users according to their need for different types of explanations.
Dupree et al.~\cite{dupree2016privacy} created personas for privacy and security practices. They considered knowledge about privacy and security and the effort users are willing to put in to protect security and privacy. 
In contrast, we considered distinct factors such as experience in dairy farming and farm size, as these factors are unique to our specific use case. Our personas illustrate users' diverse explanation needs in the dairy farming sector and enable requirements engineers to empathize with different types of farmers.

\section{Conclusions and Outlook}
\label{sec:conclusion}

In our work, we explored the explainability requirements of dairy farmers by creating user personas to gain insights into their specific needs for explanation.
In addition, we analyzed correlations between sociodemographic factors and explainability requirements.
To this end, we defined three scenarios that reflect everyday situations faced by dairy farmers.

Based on quantitative and qualitative data, we identified five distinct dairy farmer personas with varying requirements for technical and data privacy-related explanations.
Three personas (the Carefree Power-User, the Pragmatist, and the Minimalist) prefer low to moderate detail on data privacy.
Their requirements for technical explanations vary widely, from minimal to very detailed.
Together, they represent half of the participants.
The remaining two personas (the Detail-Oriented Privacy-Skeptic and the Selective Privacy-Skeptic) require highly detailed data privacy explanations.
They differ in technical requirements: the former prefers extensive detail, while the latter seeks information only in specific areas.
These two personas represent the other half of the participants.
Our findings partly contradict studies such as \citet{chazette2020explainability, nunes2017systematic} suggesting that explanations should be concise, as otherwise they could overwhelm most users.

The correlation analysis revealed several significant links between sociodemographic factors and explainability requirements.
Previous research has shown that factors like age, gender, and experience influence farmers’ acceptance and trust \cite{Michels2019}.
Our study supports this and identifies gender, age, confidence, frequency and duration of digital system use as the most influential factors.

Requirements regarding explanations also varied across scenarios, with different sociodemographic factors influencing preferences depending on context.
This highlights that farmers prioritize explanations differently, underscoring the value of context-aware design.
Understanding these correlations could help software providers tailor digital farming systems more effectively.

We acknowledge the limitations of our study, but also see it as a stepping stone for future research opportunities.
Validating user personas remains a challenge due to the absence of a widely accepted standard.
As a step toward addressing this, \citet{salminen2020persona} proposed the \pps, a framework with eight constructs for assessing persona quality, which can be adapted to fit the context of a given study.
In our planned future work, we aim to apply and evaluate the use of this framework to strengthen the validation of our personas.

\section*{Acknowledgments}

This work was funded by Carl Zeiss Stiftung, Germany,
under the Sustainable Embedded AI project, Grant No.: P2021-02-009.
It was also funded by the Deutsche Forschungsgemeinschaft (DFG, German Research Foundation) under Grant No.: 470146331, project softXplain (2022-2025).

We would like to express our sincere gratitude to all participating farmers for their valuable time and feedback.
We used the following image generation tool to create the images of the user personas: \url{https://openai.com/index/dall-e-3}

\bibliographystyle{IEEEtranN}
{\footnotesize
\bibliography{Literatures}}

\end{document}